\documentclass[12pt,psfig]{article}
\usepackage{amssymb}
\usepackage{amsmath,amsthm}
\usepackage{amsfonts}
\usepackage{graphicx}
\usepackage{graphics}
\usepackage{indentfirst}
\numberwithin{equation}{section}

\usepackage{hyperref}

\begin{document}
\begin{center}\Large\textbf{Pair Creation of Open Strings between
the Unstable-Dressed D$p$-branes}
\end{center}
\vspace{0.75cm}
\begin{center}{\large Niloufar Barghi-Janyar and \large Davoud
Kamani}
\end{center}
\begin{center}
\textsl{\small{Department of Physics, Amirkabir University of
Technology (Tehran Polytechnic) \\
P.O.Box: 15875-4413, Tehran, Iran \\
e-mails: niloufarbarqi@aut.ac.ir , kamani@aut.ac.ir \\}}
\end{center}
\vspace{0.5cm}

\begin{abstract}

We study the open string pair production
via the interaction of two parallel D$p$-branes.
The branes have been dressed by the Kalb-Ramond
field, a specific $U(1)$ gauge potential and a tachyon
field of the open string spectrum. We shall
apply the boundary state formalism, in the
context of the bosonic string theory.

\end{abstract}

\textsl{Keywords}: Background fields; Boundary state;
Interaction amplitude; String pair creation.

\newpage
\section{Introduction}

One of the remarkable windows into the
string theory is the D-branes \cite{1}-\cite{4}.
The boundary state formalism is a trusty tool for
calculating the interaction between them \cite{5}-\cite{7}.
The D-branes interaction in the presence of the
background fields has interesting
properties \cite{8}-\cite{29}.
One of them is the collapse of
the branes through the open string pair creation
in close distances \cite{30}-\cite{36}. Therefore,
fundamental string formation in the interaction of
the branes at close distances is a justification for the
attractive force between the branes \cite{37}-\cite{40}.

In fact, the tachyon field as a background field on
the worldvolume of a D-brane drastically
leads to instability and
collapse of the D-brane. Thus, the brane collapses through
the tachyon condensation into an unstable lower
dimensional brane
\cite{41}-\cite{49}. The intermediate brane
eventually collapses to the closed string vacuum or
decays to a lower dimensional stable brane
\cite{50}-\cite{56}.
Not only presence of the tachyon field as a background
field leads to instability,
but also the presence of the other
background fields, e.g. the electric and magnetic fluxes,
also gives rise to a tachyonic mode
\cite{57}, \cite{58}. In the interaction of the branes,
when the branes separation is small,
this tachyonic mode also eventuates to a tachyonic
instability \cite{32}-\cite{35}. Besides, by considering the
tachyon condensation, a system of the
separated D$p$-brane and anti-D$p$-brane
has been also investigated \cite{72}.

In this paper, at first, we introduce the interaction
amplitude of two parallel D$p$-branes
through the closed string exchange. Each brane has
been dressed by the Kalb-Ramond field,
the $U(1)$ gauge potential in a special
gauge and an open string tachyon field.
Then, we rewrite the interaction amplitude
in terms of the 1-loop annulus diagram of open string.
We shall see that
the latter from of the amplitude obviously
reveals the decay of the system via the open string
pair production. Simultaneous presence of the tachyon field,
electric and magnetic fluxes dedicates an adjustable form
and a generalized feature to the production
rate of the open strings.

This paper is organized as follows. In Sec. 2, we
shall introduce the interaction amplitude
between two parallel dressed-unstable
D$p$-branes via the boundary state formalism.
In Sec. 3, the production rate of the
open string pairs will be computed.
Section 4 is devoted to the conclusions.

\section{The interaction amplitude}

For obtaining the boundary state, associated with the
D$p$-brane, we start with the
following sigma-model action for the closed string
\cite{43}, \cite{44}, \cite{50}, \cite{59}, \cite{64}-\cite{71},
\begin{eqnarray}
S=&-&\frac{1}{4\pi\alpha'}\int_{\Sigma} {\rm d}^2\sigma
\left(\sqrt{-h} h^{ab}G_{\mu\nu} \partial_{a}X^\mu
\partial_b X^{\nu} +\varepsilon^{ab}
B_{\mu\nu}\partial_a X^{\mu}
\partial_b X^{\nu}\right)
\nonumber\\[10pt]
&+&\frac{1}{2\pi\alpha'}\int_{\partial\Sigma}
{\rm d}\sigma\left(A_{\alpha}
\partial_{\sigma}X^{\alpha}
+T^2(X^\alpha) \right),
\label{eq:2.1}
\end{eqnarray}
where $G_{\mu\nu}$, $B_{\mu\nu}$, $A_{\alpha}$,
$T^{2}(X)$, $\Sigma$ and $\partial{\Sigma}$ are the
spacetime metric, Kalb-Ramond
field, $U(1)$ gauge potential, open string
tachyon field, the worldsheet of a closed string, and the
boundary of the worldsheet, respectively.
The indices $\alpha,\beta \in \{0,1,\cdots,p\}$
label the brane worldvolume directions.
We apply the flat worldsheet
$h_{\alpha\beta}=\eta_{\alpha\beta}=\rm{diag}{(-1,1)}$,
which has been embedded in the flat spacetime
$G_{\mu\nu}=\eta_{\mu\nu}=\rm{diag}{(-1,1,...)}$.
Besides, the Kalb-Ramond field $B_{\mu\nu}$ will be
constant.

In fact, the boundary action, under an arbitrary
gauge transformation of the field
$A_{\alpha}$ is invariant. Thus,
for the $U(1)$ gauge potential, which
lives in the worldvolume of the brane, we select the
the Landau gauge
$A_{\alpha}=-\frac{1}{2}F_{\alpha\beta}X^{\beta}$
with the constant field strength
$F_{\alpha\beta}$. In addition, for the
tachyon profile we use the quadratic form
$T^2=\frac{i}{2}U_{\alpha\beta}X^{\alpha}X^{\beta}$
in which the symmetric matrix $U_{\alpha\beta}$ is
constant. Note that the gauge potential and tachyon field
belong to the open string spectrum.

\subsection{The associated boundary state to the D$p$-brane}

Vanishing of the variation of the action (\ref{eq:2.1})
gives the equation of motion and the following
boundary state equations
\begin{eqnarray}
&~&\left( \eta_{\alpha\beta} \partial_\tau X^\beta
+\mathcal{F}_{\alpha\beta}
\partial_{\sigma} X^\beta+i U_{\alpha\beta}
X^\beta \right)_{\tau=0} |B_x\rangle=0,
\nonumber\\[7pt]
&~&\left(X^i-y^i\right)_{\tau=0}|B_x\rangle=0,
\label{e:2.2}
\end{eqnarray}
where the total field strength is
$\mathcal{F}_{\alpha\beta}=F_{\alpha\beta}
-B_{\alpha\beta}$, and the parameters
$\{y^i| i= p+1,\cdots, d-1\}$
indicate the brane position.

Note that we obtained these boundary conditions
without using of the chosen gauge.
Precisely, they are valid for arbitrary gauge potential.
Since the field
strength $F_{\alpha\beta}$ is gauge invariant,
the boundary state $|B_x \rangle$
also is gauge symmetric.
However, for an arbitrary gauge field,
the field strength $F_{\alpha\beta}$
is prominently depended
to the closed string coordinates
$X^\alpha(\sigma,\tau)$.
Hence, the first equation of (\ref{e:2.2})
becomes very complicated.
For receiving a solvable equation, we employ
a constant field strength \cite{13},
\cite{60}, \cite{61}, \cite{73}-\cite{78}. Therefore,
the Landau gauge obviously enables us to
solve Eqs. (\ref{e:2.2}).

Now consider the decomposition
$|B_x\rangle = |B\rangle^{(\rm osc)}\otimes |B\rangle^{(0)}$.
Introducing the closed string mode expansion into
Eqs. (\ref{e:2.2}), the boundary state equations
are rewritten in terms of the string oscillators
and zero modes
\begin{eqnarray}
&~&\left[\left(\eta_{\alpha\beta}-\mathcal{F}_{\alpha\beta}
-\frac{1}{2m}U_{\alpha\beta}\right)\alpha^\beta_m
+\left(\eta_{\alpha\beta}+\mathcal{F}_{\alpha\beta}
+\frac{1}{2m}U_{\alpha\beta}\right)
\tilde{\alpha}^\beta_{-m}\right]
|B\rangle^{({\rm osc})}=0,
\nonumber\\[10pt]
&~&\left(\alpha^i_m-\tilde{\alpha}^i_{-m}\right)
|B\rangle^{({\rm osc})}=0,
\end{eqnarray}
for the oscillating part, and
\begin{eqnarray}
&~&\left(2 \alpha'\eta_{\alpha\beta}
p^\beta+i U_{\alpha\beta} x^\beta\right)|B\rangle^{(0)}=0,
\nonumber\\[10pt]
&~&\left(x^i-y^i\right)|B\rangle^{(0)}=0,
\label{e.2-5}
\end{eqnarray}
for the zero-mode part.

By applying the techniques of quantum mechanics, specially
the coherent state method \cite{63}, the
oscillating part of the boundary state
$|B\rangle^{(\rm osc)}$ is obtained as in the following
\begin{eqnarray}
|B\rangle^{(\rm osc)} &=&
\prod_{m=1}^{\infty} [\det M_{(m)}]^{-1}
\exp\left[ -\sum_{n=1}^{\infty}
\left( \frac{1}{n} \alpha^{\mu}_{-n}
S_{(n)\mu\nu} \tilde{\alpha}^\nu _{-n} \right) \right]
|0\rangle_\alpha |0\rangle_{\tilde \alpha},
\label{e:2.5}
\end{eqnarray}
where the matrices $S_{(m)}$, $M_{(m)}$,
$N_{(m)}$ and $Q_{(m)}$ are defined by
\begin{eqnarray}
&~&S_{(m)\mu\nu}=\left(Q_{(m)\alpha\beta}=
(M_{(m)}^{-1}N_{(m)})_{\alpha\beta}\;,\;-\delta_{i j}\right),
\nonumber\\[10pt]
&~&M_{(m)\alpha\beta}= \eta_{\alpha\beta}
-\mathcal{F}_{\alpha\beta}-\frac{1}{2m}
U_{\alpha\beta},
\nonumber\\[10pt]
&~&N_{(m)\alpha\beta}= \eta_{\alpha\beta}+
\mathcal{F}_{\alpha\beta}+\frac{1}{2m}
U_{\alpha\beta}.
\label{e:2.6}
\end{eqnarray}

By employing the quantum mechanical commutation
relation between $x^\alpha$ and $p^\beta$,
the zero-mode part of the boundary state
$|B\rangle^{(0)}$ possesses the following feature
\begin{eqnarray}
|B\rangle^{(0)}&=&\frac{T_p}{2\sqrt{\det(U/4\pi \alpha')}}
\; \delta^{(d-p-1)}(x^i-y^i) |p^i=0\rangle
\nonumber\\
&\times& \int_{-\infty}^{\infty} \prod^p_{\alpha=0}
\exp\left[2\alpha' \sum_{\beta\neq\alpha}
(U^{-1})_{\alpha\beta}p^\alpha p^\beta
+\alpha'(U^{-1})_{\alpha\alpha}(p^{\alpha})^{2}\right]
|p^\alpha \rangle {\rm d}p^\alpha,
\label{e:2.7}
\end{eqnarray}
where $T_{p}$ is the tension of the D$p$-brane.
For obtaining the normalization prefactors
of Eqs. (\ref{e:2.5}) and (\ref{e:2.7}), via
the Gaussian path integral, we employed the Landau gauge
$A_{\alpha}=-\frac{1}{2}F_{\alpha\beta}X^{\beta}$
and the quadratic form of the tachyon profile. These
known choices dedicate a quadratic
form to the boundary action of
Eq. (\ref{eq:2.1}). Consequently, the normalization prefactors
have been computed from
the disk partition function. Similar normalization
factors can be found, e.g., in the Refs. \cite{60}-\cite{62}.

The total boundary state is given by
\begin{equation}
|B\rangle=|B\rangle^{(\rm osc)} \otimes |B\rangle^{(0)}
\otimes |B\rangle^{(\rm gh)},
\label{e:2.8}
\end{equation}
where $|B\rangle^{\rm (gh)}$ is the boundary state
of the conformal ghosts.
Since the ghosts fields do not interact with the matter
part, their contribution to the boundary state is not
influenced by the background fields. Thus,
$|B\rangle^{\rm (gh)}$ is given by
\begin{equation}
|B\rangle^{\rm (gh)}=\exp \left[ \sum_{n=1}^\infty
(c_{-n}\tilde b_{-n}-b_{-n} \tilde c_{-n})\right]
\frac{c_0+\tilde c_0}{2}\; |q=1\rangle\; |\tilde q=1\rangle.
\label{e:2.9}
\end{equation}

\subsection{The interaction of two D$p$-branes}

For obtaining the open string pair creation, we need
to calculate the interaction amplitude of the branes.
The interaction can take place via the closed string
exchange. Hence, the amplitude is given by
\begin{eqnarray}
\mathcal{A}_{\rm{closed}} & = & \langle B_1|D|B_2\rangle ,
\nonumber\\
D & = & 2\alpha' \int^\infty_0 {\rm d}t\;e^{-tH},
\end{eqnarray}
where ``$D$'' is the closed string propagator, and
``$H$'' represents the total closed string Hamiltonian.
The Hamiltonian contains the matter part and
ghost portion, as in the following
\begin{eqnarray}
H &=& \alpha'p^{\mu}p_{\mu}+
2 \sum^\infty_{n=1}\left(\alpha^\mu_{-n}\alpha_{n\mu}+
\tilde{\alpha}^\mu_{-n}\tilde{\alpha}_{n\mu}\right)
\nonumber\\
&+&\sum^\infty_{n=1}n\left(b_{-n}c_n
+c_{-n}b_n+\tilde{b}_{-n}\tilde{c}_n+
\tilde{c}_{-n}\tilde{b}_n\right)-\frac{35}{12}.
\end{eqnarray}
Introducing Eqs. (\ref{e:2.5}) and (\ref{e:2.7})
into this amplitude we receive
\begin{eqnarray}
\mathcal{A}_{\rm{closed}}
&=&-(-2)^{-([(p-1)/2]+3)/2}\pi^{(p+5)/2}\;T^2_p\;V_{p+1}
\frac{\prod_{m=1}^\infty \left[\det
\left(M_{(m)1}^\dagger M_{(m)2}
\right) \right]^{-1}}{\sqrt{\det(U_1/{4\pi\alpha'})
\det(U_2/{4\pi\alpha'})}}\;
\nonumber\\[10pt]
&\times& \int_0^\infty{\rm d}t\bigg\{
\exp{\left(-\frac{y^2}{2\pi\alpha' t}\right)}
\left(\det{\left[\frac{\alpha'\pi t}{2}\mathbb{I}
-2\alpha'\left(U_{1}^{-1}+U_{2}^{-1}\right)
\right]}\right)^{-1/2}
\nonumber\\[10pt]
&\times&\left(\frac{1}{2\pi^{2}\alpha' t}
\right)^{(d-p-1)/2} |z|^{(d-p-25)/12}
\left(\eta(it) \right)^{p-d+4}
\prod_{m=1}^{\infty}\prod_{a=0}^{\left[\frac{p-1}{2}\right]}
\frac{\sin{(\pi\nu_{(a,m)}})}{\Theta_{1}
(\nu_{(a,m)}|i t)}\bigg\},
\label{e:2.12}
\end{eqnarray}
where $y^2=\sum_{i=p+1}^{d-1}(y_{1}^{i}
-y_{2}^{i})^{2}$ indicates the square
distance between the branes,
and $V_{p+1}$ represents the worldvolume of each
brane. The second line of this amplitude
is the contribution of the zero-mode parts of the boundary
states. The exponential factor in this line
clarifies that by increasing the distance between
the branes, their interaction exponentially is damped.
The last line of Eq. (\ref{e:2.12}) represents the
effects of the oscillating portions of the boundary states.
The $\eta$-function originates from the
oscillators of the perpendicular directions
and the conformal ghosts.
The quantity $[(p-1)/2]$ is the integer part of
$(p - 1)/2$. Besides, we used the theta- and eta-functions
\begin{eqnarray}
\Theta_{1}\bigg(\nu_{(a,m)}|i t\bigg)
&=&-2 |z|^{1/4}\sin{(\pi\nu_{(a,m)})}
\nonumber\\
&\times& \prod_{n=1}^{\infty}
\left[\left(1-|z|^{2n}\right)
\left(1-e^{2i\pi \nu_{(a,m)}}|z|^{2n}\right)
\left(1-e^{-2i\pi\nu_{(a,m)} }|z|^{2n}\right)\right],
\nonumber\\[10pt]
\eta(i t)&=&|z|^{1/12}
\prod_{n=1}^{\infty}\left(1-|z|^{2n}\right),
\end{eqnarray}
where $|z|=e^{-\pi t}$. The constants
$\lambda_{(a,m)} \equiv e^{2\pi i\nu_{(a,m)}}$
and $\lambda_{(a,m)}^{-1}$ are the eigenvalues
of the $(p+1)\times(p+1)$ orthogonal matrix
$\Lambda_{(m)}=Q_{(m)2}Q^{\rm T}_{(m)1}$.
Note that a real orthogonal matrix
possesses pairs of eigenvalues with the
modulus one, and/or plus/minus one eigenvalues.
In each of the pairs, one eigenvalue is the
complex conjugate of the other.

By adjusting the parameters of the setup,
the matrix $\Lambda_{(m)}$ becomes orthogonal.
In this case, the eigenvalues of $\Lambda_{(m)}$
for the odd (even) values of $p$ are
labeled by the numbers $a=0,1,2,\cdots, [(p-1)/2]$.
Besides, for the even values of $p$
there is one additional unit eigenvalue,
i.e. $\lambda_m=1$.
The structure of the matrix $\Lambda_{(m)}$
implies that the eigenvalues manifestly depend on the
parameters of the configuration.
Therefore, the case $\nu_{(a,m)}= 0$,
with all possible values of $a$ and $m$,
is corresponding to the setup
without the background fields.
We shall see that the orthogonality of
$\Lambda_{(m)}$ specifies that at least one of the
$\nu_{(a,m)}$s should be purely imaginary.

When all $\nu_{(a,m)}$s are real the amplitude
(\ref{e:2.12}) is valid for any distance of the branes.
The factor $\exp{(-y^2/2\pi\alpha' t)}$
implies that for the small distances of the branes,
the small $t$-integration becomes important.
If we consider one of the
$\nu_{(a,m)}$s, e.g. $\nu_{(0,m)}$,
to be a pure imaginary $\nu_{(0,m)}=i\upsilon_{(0,m)}$,
then we receive $\upsilon_{(0,m)}\in{(0,\infty)}$.
Thus, for the small values of $t$, the factor
$1 +|z|^{4n} -2|z|^{2n}\cosh(2\pi\upsilon_{(0,m)})
\approx 2(1-\cosh(2\pi\upsilon_{(0,m)}))$
in the denominator of the infinite product
in Eq. (\ref{e:2.12})
becomes negative. Since the infinite
product contains an infinite number of such factors,
the sign of the amplitude obviously is ambiguous.
This ambiguity potentially exhibits the occurrence
of a new phenomenon, which refers to the decay of
the underlying system via the open
string pair creation \cite{32}-\cite{35}.

We saw that all $\nu_{(a,m)}$s depended on the background
fields, i.e. the electric, magnetic and tachyon
fields. This implies that by adjusting the
values of the background fields
one can conveniently receive a pure imaginary
set $\{\nu_{(0,m)}| m \in \mathbb{N}\}$.

We should note that the determinant in the second
line of Eq. (\ref{e:2.12}) is
the characteristic polynomial
of the real and symmetric matrix
$R=4(U^{-1}_{1}+U^{-1}_2)/\pi$. Thus, the matrix $R$
can be diagonalized with the real eigenvalues.
To have positive $\rm{det}(t.\mathbb{I}-R)$ for
any positive ``$t$'', the eigenvalues of the odd
multiplicity have to be negative.
Besides, in order to avoid any singularity
in the integral, since the inverse of
the square root of the determinant has been
appeared, the eigenvalues with the even
multiplicity should also be negative. Hence, all
eigenvalues have to be negative. The negativity
of the eigenvalues of the matrix $R$
restricts the parameters of the setup.
Under the above circumstances,
the foregoing determinant does not change
the sign. We shall apply this prominent
fact to extract the pair creation
rate of the open strings.

\section{Open string pair creation}

\subsection{ The 1-loop open string amplitude}

As we said we can consider all $\nu_{(0,m)}$s to be pure
imaginary quantities, i.e.
$\nu_{(0,m)}=i\upsilon_{(0,m)}$, \cite{30}.
Afterward, the closed-string cylinder
amplitude will be obscure for
small separation of the branes.
That is, due to the small value of $t$, one
receives a new phenomenon. Thus, the best way for understanding
and describing the system, is going to the open string
annulus amplitude. The latter amplitude reveals
many interesting
properties of the system such as its instabilities.
The exchange of the amplitudes will be done by
applying the  Jacobi transformation $t\to1/t$ on the cylinder
amplitude (\ref{e:2.12}),

\begin{eqnarray}
\mathcal{A}_{\rm{open}}
&=& \pi^{(p+5)/2}(2\pi^{2}\alpha')^{-(d-p-1)/2}\;
T_{p}^{2}V_{p+1}\;
\frac{\prod_{m=1}^\infty \left[\det
\left(M_{(m)1}^\dagger M_{(m)2}
\right) \right]^{-1}}{\sqrt{\det (U_1/4\pi\alpha')
\det (U_2/4\pi\alpha')}}
\nonumber\\[10pt]
&\times& \int_0^\infty{\rm d}t\bigg\{
\exp{\left(-\frac{y^2 t}{2\pi\alpha' }\right)}
\bigg(\det{\left[\frac{\alpha'\pi}{2t}\mathbb{I}
-2\alpha'(U_{1}^{-1}+U_{2}^{-1})\right]}\bigg)^{-1/2}
\nonumber\\[10pt]
&\times&e^{-([(p-1)/2]+1)\pi t/8}\;e^{-(d-p-25)\pi/12 t}
\;t^{3/2+[(p-1)/2]/4}
\nonumber\\[10pt]
&\times&\prod_{m=1}^{\infty}
\bigg[e^{-\pi\upsilon_{(0,m)}^{2}t}\;
\frac{\sinh{(\pi\upsilon_{(0,m)})}}
{\sin{(\pi\upsilon_{(0,m)} t})}
\mathcal{Z}_{m}
\prod_{a=1}^{\left[\frac{p-1}{2}\right]}
\frac{e^{\pi\nu_{(a,m)}^{2}t}
\sin{(\pi\nu_{(a,m)})}}{\sinh{(\pi\nu_{(a,m)}t)}}
\bigg]\bigg\},
\label{e:3.1}
\end{eqnarray}
where $\mathcal{Z}_{m}$ is given by
\begin{eqnarray}
\mathcal{Z}_{m}
&=&\prod^\infty_{n=1}\bigg(\left(1-|z|^{2n}\right)^{d-p-3}
\left[1-2|z|^{2n}\cos{(2\pi\upsilon_{(0,m)}t)}+|z|^{4n}
\right]
\nonumber\\[10pt]
&\times&\prod_{a=1}^{[\frac{p-1}{2}]}\left[1-2|z|^{2n}
\cosh{(2\pi\nu_{(a,m)}t)}+|z|^{4n}\right]\bigg)^{-1}\;.
\label{e:3.2}
\end{eqnarray}
For obtaining this amplitude, we applied the following
properties of the $\Theta_{1}$-function and the Dedekind
$\eta$-function
\begin{eqnarray}
\eta(\tau)=\frac{1}{\sqrt{-i\tau}}\;\eta \left(-\frac{1}{\tau}
\right),
\hspace{0.6cm}\Theta_{1}(\nu_{(a,n)}|\tau)
=i \frac{e^{-i \pi \nu_{(a,n)}^{2}/\tau}}{\sqrt{-i\tau}}\;
\Theta_{1}\left(\frac{\nu_{(a,n)}}{\tau}
{\bigg |}-\frac{1}{\tau}\right),
\label{e:3.3}
\end{eqnarray}
where the time variable is  $\tau = i\times t$.

\subsection{Pair creation of the open strings}

Here we briefly clarify
the reason that gives rise to the instability
of our system. Performing the limit $t\to\infty$
on Eq. (\ref{e:3.1}), the variable exponential
$\exp{(-2\pi\alpha' M^2_{\rm eff}t)}$ is revealed,
where
\begin{eqnarray}
M^2_{\rm eff}=
\left(\frac{y}{2\pi\alpha'}\right)^2
+\frac{\left[(p-1)/2\right]+1}{16\alpha'}
-\frac{1}{2\pi\alpha'}
\sum_{m=1}^{\infty}\left[
\sum_{a=1}^{\left[\frac{p-1}{2}\right]}\nu_{(a,m)}
(\nu_{(a,m)}-1)-\upsilon^{2}_{(0,m)}\right].
\end{eqnarray}
The quantity $M^2_{\rm eff}$ defines the effective
mass square of the open string,
which connects the D$p$-branes
to each other. A phase transition,
which is called the tachyon condensation, happens when
the effective mass square becomes negative. This
induces the condition
\begin{eqnarray}
y<\pi\;\sqrt{2\alpha'
\left[\frac{1}{\pi}\sum_{m=1}^{\infty}\left(\sum_{a=1}^
{\left[\frac{p-1}{2}\right]}\nu_{(a,m)}(\nu_{(a,m)}-1)-
\upsilon_{(0,m)}^{2}\right)-\frac{[(p-1)/2]+1}{8}\right]}\;.
\end{eqnarray}
The square root also establishes a new condition, i.e.,
the phrase under the square root should be
positive. These conditions demonstrate that the
appearance of the tachyonic mode prominently
depends on the values of the parameters of the setup.
However, when the foregoing conditions
are satisfied the system drastically undergoes
the tachyonic instability, and hence the
integrand of Eq. (\ref{e:3.1})
blows up for the large values of $t$.

The imaginary part of the
interaction amplitude defines the open string
pairs creation.
The simple poles in Eq. (\ref{e:3.1})
refer to the existence of an
imaginary part for the amplitude.
The factor $\sin{(\pi\upsilon_{(0,m)}t)}$
gives an infinite number of simple poles
along the positive $t$-axis, i.e.
$t_{(k,m)} = k/\upsilon_{(0,m)}\ge1$ with $k \ge 1$.
Each of these poles individually
indicates the production of an open string pair
and the decay of the system. Finally,
we obtain the following decay rate
per unit volume of the $Dp$-brane
through the open string pair creation
\begin{eqnarray}
\mathcal{W}_{p, p}
&=&-\frac{2 \rm{Im}\mathcal{A}_{\rm{open}}}{V_{p+1}}
\nonumber\\
&=& \pi^{(7-2d+4p)/2}(2\alpha')^{(1-d+p)/2}\;T^2_p\;
\frac{\prod_{m=1}^{\infty}\left[\det\left(M_{(m)1}^\dagger
M_{(m)2}\right) \right]^{-1}}
{\sqrt{\det (U_1/4\pi\alpha')\det (U_2/4\pi\alpha')}}
\nonumber\\
&\times&\sum_{k=1}^{\infty}(-1)^{k+1}\bigg\{
\sum_{m=1}^{\infty}\bigg[
\left(k/\upsilon_{(0,m)}\right)^{3/2+[(p-1)/2]/4}
\nonumber\\
&\times&\exp{\left(-\frac{y^{2} k}{2\pi\alpha'
\upsilon_{(0,m)}}-\frac{(d-p-25)\pi\upsilon_{(0,m)}}
{12 k}-\frac{[(p-1)/2]\pi k}{8\upsilon_{(0,m)}}-
\pi k\upsilon_{(0,m)}\right)}
\nonumber\\
&\times&\left(\det\left[\frac{\alpha'\pi\upsilon_{(0,m)}}
{2k}\mathbb{I}-2\alpha'\left(U_{1}^{-1}+U_{2}^{-1}
\right)\right]\right)^{-1/2}
\frac{\sinh{(\pi\upsilon_{(0,m)})}}
{\prod^\infty_{n=1}\left(1-
e^{-4 n k \pi/\upsilon_{(0,m)}}\right)^{d-p-1}}
\nonumber\\
&\times&\prod_{a=1}^{\left[\frac{p-1}{2}\right]}\left(
\frac{e^{\pi k \nu_{(a,m)}^{2}/\upsilon_{(0,m)}}
\sin{(\pi\nu_{(a,m)})}}
{\prod^\infty_{n=1}\left(1-2e^{-2 n k\pi/\upsilon_{(0,m)}}
\cosh{\left(\frac{2 k\pi\nu_{(a,m)}}
{\upsilon_{(0,m)}}\right)}+e^{-4 n k \pi/\upsilon_{(0,m)}
}\right)}\right)\bigg]\bigg\}.
\label{e:3.6}
\end{eqnarray}
Since all $\nu_{(a,m)}$s and $\upsilon_{(0,m)}$s
depend on the background fields the decay rate
$\mathcal{W}_{p, p}$ is a complicated
function of the parameters of the setup.
We observed that the mass of a created open string
is proportional to the brane separation.
This implies that in the large distance of the branes
the probability of the open
string pair creation is small, as expected.
However, for the finite distance, specially
when the branes are near to each other, we receive the
open string pair creation and subsequently
the decay of the system. Note that the production
of open strings between the branes has a
resemblance with the Casimir effect.

In fact, the presence of the
electric fields obviously plays the main role in the
production of the open string pairs. Besides, the
appearance of the magnetic fluxes
give rise to the open string tachyonic instability
when the brane separation sufficiently is small.
In addition, the presence of magnetic
background usually causes
an enhancement of the rate of the open string pair
creation (e.g. see Refs. \cite{30}-\cite{35}).
We observed that the tachyon fields
also drastically influenced the pair creation rate.

\section{Conclusions}

In the beginning, we introduced
a boundary state which is
corresponding to a dressed-unstable
D$p$-brane in the presence of
a constant Kalb-Ramond field,
an internal $U(1)$ gauge potential
in a specific gauge and a quadratic
open string tachyon field.
Then the interaction amplitude for two
parallel D$p$-branes, in the presence of the
foregoing fields, was introduced.
For extracting the open string pair creation
we converted the cylinder amplitude of closed string to
the 1-loop amplitude of open string.
This exchange conveniently was acquired by
applying the Jacobian transformation on the
former amplitude.

The behavior of the open string
amplitude for the large values of the integration
variable ``$t$'' was studied.
Due to the tachyon condensation,
the instability of the system was created. This instability
is specified by two conditions on the parameters
of the configuration.

The open string annulus amplitude enabled us
to extract the rate of the open string production.
This pair creation and the decay of the system
extremely depend on the electric, magnetic and tachyonic
fields on the interacting branes.
The new parameters, i.e. the tachyons matrix elements
$U_{(1)\alpha\beta}$ and $U_{(2)\alpha\beta}$,
possessed a remarkable effect on the pair creation rate.
Finally, by varying the parameters of the setup
the value of the production rate can
be accurately adjusted to any desirable value.


\end{document}